\documentclass[twocolumn,twocolappendix]{aastex631}
\usepackage{comment}
\usepackage{chemformula}
\usepackage{booktabs}
\usepackage{threeparttable}
\usepackage{rotating}
\usepackage{mathtools}
\usepackage{graphicx}
\usepackage{soul}
\usepackage{hyperref}
\usepackage{natbib}

\graphicspath{{./}{figures/}}

\received{January 1, 2018}
\revised{January 7, 2018}
\accepted{\today}
\submitjournal{ApJ}

\shortauthors{Ferrero et al.}

\begin{document}

\title{Formation of complex organic molecules on interstellar CO ices? Insights from computational chemistry simulations}

\correspondingauthor{Albert Rimola, Mariona Sodupe}
\email{albert.rimola@uab.cat, mariona.sodupe@uab.cat}

\author[0000-0001-7819-7657]{Stefano Ferrero}
\affiliation{Departament de Qu\'{i}mica, Universitat Aut\`{o}noma de Barcelona, Bellaterra, 08193, Catalonia, Spain}

\author[0000-0001-9664-6292]{Cecilia Ceccarelli}
\affiliation{Univ. Grenoble Alpes, CNRS, Institut de Plan\'{e}tologie et d'Astrophysique de Grenoble (IPAG), 38000 Grenoble, France}

\author[0000-0001-8886-9832]{Piero Ugliengo}
\affiliation{Dipartimento di Chimica and Nanostructured Interfaces and Surfaces (NIS) Centre, Universit\`{a} degli Studi di Torino, via P. Giuria 7, 10125, Torino, Italy}

\author[0000-0003-0276-0524]{Mariona Sodupe}
\affiliation{Departament de Qu\'{i}mica, Universitat Aut\`{o}noma de Barcelona, Bellaterra, 08193, Catalonia, Spain}

\author[0000-0002-9637-4554]{Albert Rimola}
\affiliation{Departament de Qu\'{i}mica, Universitat Aut\`{o}noma de Barcelona, Bellaterra, 08193, Catalonia, Spain}



\begin{abstract}

Carbon ($^3$P) atom is a reactive species that, according to laboratory experiments and theoretical calculations, condensates with interstellar ice components. This fact is of uttermost importance for the chemistry in the interstellar medium (ISM) because the condensation reaction is barrierless and the subsequent  species formed are still reactive given their open-shell character. Carbon condensation on CO-rich ices forms the \ch{C=C=O} ($^3$$\Sigma$$^-$) species, which can be easily hydrogenated twice to form ketene (H$_2$CCO). Ketene is very reactive in terrestrial conditions, usually found as an intermediate hard to be isolated in chemical synthesis laboratories. These characteristics suggest that ketene can be a good candidate to form interstellar complex organic molecules (iCOMs) via a two-step process, i.e., its activation followed by a radical-radical coupling. In this work, reactions between ketene and atomic H, and the OH and NH$_2$ radicals on a CO-rich ice model have been explored by means of quantum chemical calculations complemented by kinetic calculations to evaluate if they are favourable in the ISM. 
Results indicate that H addition to ketene (helped by tunneling) to form the acetyl radical (CH$_3$CO) is the most preferred path, as the reactions with OH and NH$_2$ possess activation energies ($\geq$ 9kJ/mol) hard to surmount in the ISM conditions, unless external processes provide energy to the system.
Thus, acetaldehyde (CH$_3$CHO) and, probably, ethanol (CH$_3$CH$_2$OH) formation via further hydrogenations are the possible unique operating synthetic routes. Moreover, from the computed relatively large binding energies of OH and NH$_2$ on CO ice, slow diffusion is expected, hampering possible radical-radical couplings with CH$_3$CO.
The astrophysical implications of these findings are discussed considering the incoming James Webb Space Telescope observations.


\end{abstract}

\keywords{Astrochemistry --- Interstellar medium --- Interstellar molecules --- Interstellar dust --- Surface ices --- Complex organic molecules --- Reactions rates --- Computational methods}


\section{Introduction} \label{sec:intro}

Interstellar grains are submicron-sized solid particles made either of carbonaceous materials or  silicates. 
In cold ($\sim$10 K) and dense ($\sim$10$^4$ cm$^{-3}$) molecular clouds, these grains are covered predominantly by water icy mantles, with several other, less abundant, species detected by infra-red (IR) observations of the interstellar ices \citep[e.g.,][]{boogert_observations_2015,Yang2022-JWSTices,McClure2023NatAs}. 
Interstellar grains are important in astrochemistry because they can provide the surfaces on which chemical reactions can occur forming stable products \citep[e.g.,][]{Tielens1982,cuppen_grain_2017,McCoustra-Rev,ceccarelli_organic_2022}. 

In addition to H$_2$O, one of the most abundant icy species is carbon monoxide (CO), which is thought to form in the gas phase and then to freeze out onto the surface of the grains \citep{caselli_co_1999, bacmann_degree_2002, favre_significantly_2013}. 
The CO freeze-out is supposed to happen after the formation of the bulk of water ice and, therefore, in the extreme cases of large CO freeze-out, interstellar ices are thought to present an (almost) onion-like structure, the innermost layers being formed by a polar phase dominated by water, whereas the outer layers by a non-polar phase, possibly dominated by CO \citep{boogert_observations_2015, pontoppidan_c2d_2008}. 
These non-polar outermost layers are thought to be crucial for the formation of interstellar complex organic molecules (iCOMs) through hydrogenation of CO followed by the formation of radicals via photodissociation on the ice-surfaces, and their recombination \citep{Garrod2006,chuang_production_2017, chuang_formation_2021, simons_formation_2020}. 

However, another promising route towards chemical complexity in the interstellar medium (ISM), which has emerged in the past few years, is the reactivity of atomic carbon towards different components of the icy mantles. The condensation of atomic carbon on water ice, both in its neutral (C) and in its cationic (C$^+$) forms, has been studied, and possible chemical reactions with different icy components have been elucidated \citep{krasnokutski_low-temperature_2017,shimonishi_adsorption_2018,henning_experimental_2019,woon_quantum_2021,molpeceres_carbon_2021, potapov_new_2021}.
Moreover, some of these processes have been linked to chemical pathways to form amino acids \citep{krasnokutski_condensation_2020, krasnokutski_pathway_2022}. 

On the other hand, the condensation of atomic carbon on pure CO ice is by far less studied. The C($^3$P) + CO reaction has been reported and studied in the gas phase by a recent computational work, which found the formation of the C=C=O ($^3$$\Sigma$$^-$) species as the product in a barrierless way \citep{papakondylis_electronic_2019}. 
In the ISM, these species can be easily hydrogenated twice to form ketene (H$_2$CCO), which can be hydrogenated even further to form acetaldehyde, as found recently experimentally \citep{fedoseev_hydrogenation_2022} and, eventually, ethanol. 
The formation of the latter species is particularly interesting as, in the gas phase, it is the starting point of a chain of reactions leading to glycolaldehyde \citep{Skouteris2018-ethanoltree}.
Recent observations by the James Webb Space Telescope (JWST) towards background stars have shown the possible presence of acetaldehyde and ethanol in the icy grain mantles in molecular clouds \citep{Yang2022-JWSTices,McClure2023NatAs}.
In these regions, the ices are composed of about 25\% of frozen CO (i.e., the catastrophically freeze-out mentioned above has not occurred yet) \citep[e.g.][]{boogert_observations_2015} while some atomic carbon is still in the gas phase \citep{Zmuidzinas1988,Kamegai2003}.
Although the identification of iced acetaldehyde and ethanol needs to be confirmed, the possibility that (either of) these species are already present in the molecular cloud phase warrants a dedicated study on the chemistry triggered by the C condensation on CO ice.


In this work, the formation and reactivity of ketene on a model of CO ice are explored through quantum mechanical simulations. 
Its reactivity with abundant interstellar radicals, like H, O, N, NH, OH and NH$_2$, is studied in order to identify whether and which of these species can react with ketene, thereby opening up chemical pathways that form even more complex species.

\section{Computational details}
\subsection{Gas phase calculations}
A preliminary gas phase screening of reaction barriers was made to determine which reactions are more probable at ISM conditions. All the electronic structure calculations have been carried out with the Orca 5.0 software \citep{neese_orca_2020}. Density functional theory (DFT) was used for geometry optimizations adopting the $\omega$B97X-D4 functional and the def2-TZVP basis set \citep{najibi_dft-d4_2020,weigend_balanced_2005}. (hereafter referred to as $\omega$B97X-D4/TZVP). Geometry optimizations were carried out with the geometrical counterpoise correction (gCP) method to remove the basis set superposition error (BSSE) \citep{kruse_geometrical_2012, liu_accurate_1973}. Electronic energies were refined with single point calculations at the DFT optimized geometries with the CCSD(T)-F12 and DLPNO-CCSD(T)-F12 methods \citep{hattig_explicitly_2012,adler_simple_2007, kong_explicitly_2012, pavosevic_sparsemapssystematic_2017}, which employs a cc-pVTZ-F12 basis set, the cc-pVTZ-F12-CABS near-complete basis set and the aug-cc-pVTZ/C fitting basis set for the resolution of identity (RI) approximation \citep{weigend_efficient_2002,peterson_systematically_2008}. For simplicity the CCSD(T)-F12/cc-pVTZ-F12//$\omega$B97X-D4/TZVP and DLPNO-CCSD(T)-F12/cc-pVTZ-F12//$\omega$B97X-D4/TZVP levels of theory will be referred to as CCSD(T)-F12 and DLPNO-F12, respectively. For DLPNO calculations, a tight PNO setting was used. Transition state (TS) structure searches have been conducted using the NEB-TS algorithm implemented in Orca \citep{asgeirsson_nudged_2021}. Harmonic vibrational frequencies were calculated to characterize the nature (e.g. minimum or TS) of the optimized structures and to correct the electronic energies for the zero point energy (ZPE).  

\subsection{Solid phase calculations} \label{sec:solid_phase}
\begin{figure*}
    \centering
    \includegraphics[width=\textwidth]{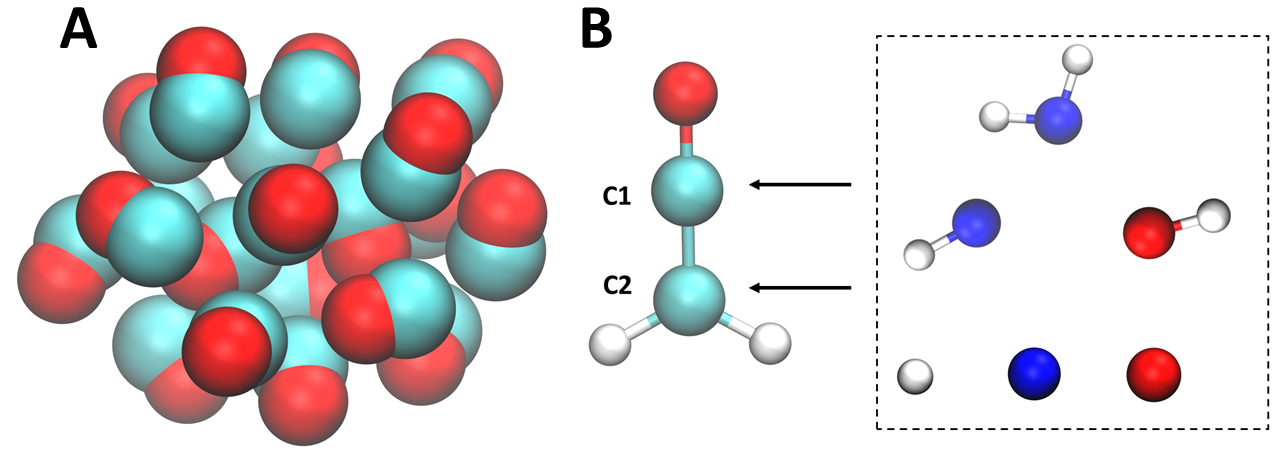}
    \caption{\textit{Panel A:} Optimized structure of the CO cluster model in Van der Waals representation. \textit{Panel B, left:} Ketene molecule structure with the C1 and C2 labels. 
    \textit{Panel B, right (dashed box)}: Molecular structure of the species to react with ketene (H, N, O, NH, OH and NH$_2$) considering the two possible attacks (arrows towards C1 and C2). Atom colour legend: H, white; C, cyan; N, blue; O, red.}
    \label{fig:figure1}
\end{figure*}

In order to mimic a CO ice, a cluster approach was used in a similar way to the inspiring paper by  \citet{lamberts_formation_2019}. 
An initial cluster was generated with the Packmol software by randomly placing 20 CO molecules inside a sphere of 12 \AA ~radius, which was then optimized with the $\omega$B97X-D4 functional and a def2-SVP basis set (see Figure \ref{fig:figure1}A). This cluster model is enough because, since the nature of the interactions between CO molecules (namely, quadrupole-quadrupole and dispersion components, \citet{Zamirri2018}), the generated cluster model exhibits all the possible CO orientations on the surface, that is, C/C, O/O and C/O, hence covering all the likely CO configurational variability.

As CO molecules in the cluster are not strongly bound together, and to avoid large deformation of the cluster structure when studying adsorption and reactivity, constrained geometry optimizations and transition state searches were performed, as explained as follows.\\ 
Interaction energies for OH and NH$_2$ on the CO cluster were calculated by performing geometry optimizations, in which only the adsorbate (NH$_2$ or OH) and the closest CO molecules (3 to 5, depending on the site) were allowed to relax. Harmonic frequencies were calculated only for the adsorbates employing a partial hessian vibrational analysis (PHVA) scheme \citep{li2002partial}. The ZPE-corrected interaction energies on the CO ice model were calculated as:

\begin{equation}
    \Delta \text{H}(0) = E_{complex} - E_{CO} - E_{adsorbate} + \Delta ZPE
\end{equation}

where $E_{complex}$ is the absolute potential energy of the adsorption complex, $E_{CO}$ is the absolute potential energy of the isolated optimized CO cluster and $E_{adsorbate}$ is the absolute potential energy of the isolated adsorbate, and bearing in mind that, at 0 K, the absolute ZPE energy is equal to the absolute enthalpy, i.e., E$_0$ = H(0). $\Delta ZPE$ values have been calculated by subtracting the ZPE corrections of the adsorbate optimized on the CO cluster and the ZPE corrections of the isolated adsorbate.

For the reactivity on the CO cluster, the structures of reactants and products were first optimized, and then TS structures were localized with the NEB-TS algorithm. Seven CO molecules were included in the optimizations and TS searches. A PHVA scheme was used to characterize the optimized structures as minimum or TS and to correct for ZPE. Energy barriers were refined by single point calculations at DLPNO-CCSD(T)-F12 level of theory and calculated as:       

\begin{equation}
\Delta \text{H}^\ddag(0) = E_{TS} - E_{min} + \Delta ZPE    
\end{equation}

where $E_{TS}$ and $E_{min}$ are the DLPNO-CCSD(T)-F12 absolute potential energies for the transition state and the minimum structure of the reaction, respectively. $\Delta ZPE$ has been calculated by subtracting the ZPE of the fragment made by the adduct (ketene plus H/OH/NH$_2$) optimized on the CO cluster and the ZPE of the adduct calculated in vacuum at $\omega$B97X-D4/TZVP level. At the ISM  temperatures (e.g., 10 K), thermal corrections are in practice negligible (\cite{Zamirri2017, Zamirri2019, enrique-romero_reactivity_2019, enrique-romero_theoretical_2021}) so we assume that the calculated $\Delta H(0)$ and $\Delta H^\ddag(0)$ do not vary at the cryogenic temperatures. The VMD software was used for rendering images \citep{humphrey_vmd_1996}.

\subsubsection{Instanton rate theory calculations} \label{sec:inst_rate_theory}
To assess the effect of tunneling on the reaction rates, the semiclassical instanton theory \citep{miller_semiclassical_1975, chapman_semiclassical_1975} was employed for the H additions to ketene. A simple estimation of the crossover temperature (T$_c$), at which tunneling becomes important, was obtained as:

\begin{equation}
T_c = \frac{\hbar \omega}{2\pi k_B }    
\end{equation}

where h is the Planck’s constant, k$_B$ is the Boltzmann’s constant and $\omega$ is the imaginary frequency of the TS. As instanton theory tends to overestimate the reaction rates around T$_c$ \citep{andersson_comparison_2009}, alongside that our interest is on rates at interstellar temperatures, instanton rate theory has been applied only in the deep tunneling regime, e.g. below T$_c$. The instanton describes the most probable tunneling path from reactants to products at a given temperature and can be regarded as a saddle point on a ring polymer potential energy surface constructed as a discretized Feynman path of N segments, called beads \citep{kastner_theory_2014, beyer_quantum_2016, richardson_perspective_2018, richardson_ring-polymer_2018}. Instantons for the hydrogenation reactions on the two ketene carbon atoms have been optimized employing a progressive cooling approach, starting from a temperature just below T$_c$ down to 50 K. As a first discretization of the path, 16 beads were employed, which were then incremented up to 128 beads to obtain convergence on the rates at 50 K. This was the ending temperature because convergence at lower temperatures requires even more beads, making the calculation computationally impractical. However, as it will be seen, the rate constants at 50 K do not depend on temperature so they can be extrapolated to 10 K. A duel-level instanton approach (CCSD(T)-F12/cc-pVTZ-F12//$\omega$B97X-D4/TZVP) \citep{meisner_dual-level_2018} was then used to refine the energy of the beads. Finally, as these reactions are supposed to happen on a CO-rich ice, the implicit surface model approach was applied to include surface effects \citep{meisner_atom_2017}, which holds only if the catalytic
role of the surface is negligible. In this approximation, the rotational partition function is assumed to be constant during the reaction as the surface hampers rotations in both the reactant and TS structures. Instantons have been optimized on the fly by interfacing the Orca software with a Python code developed in Jeremy Richardson’s group at ETH Zurich.

\section{Results and discussion}
\subsection{Gas phase calculations}
The C($^3$P) + CO reaction, already studied in the work of \citet{papakondylis_electronic_2019}, was here reproduced at the $\omega$B97x-D4/TZVP level of theory. Results are in agreement with the previous findings on the barrierless formation of \ch{C=C=O} ($^3\Sigma^-$) and, thus, the formation of ketene via a double hydrogenation reaction is viable. In order to study the reactivity of ketene and to assess the height of the reaction barriers, gas-phase calculations were carried out for the reactions with abundant interstellar radicals, involving attacks on the two carbon ketene atoms, that is:

\begin{equation} \label{C1}
\text{H}_{2}\text{CCO} + \text{X}^{\bullet} \ch{->} \text{H}_{2}\text{C}^{\bullet}\text{CXO}    
\end{equation}

\begin{equation} \label{C2}
\text{H}_{2}\text{CCO} + \text{X}^{\bullet} \ch{->} \text{H}_{2}\text{XC}\text{C}^{\bullet}\text{O}
\end{equation}

with X$^\bullet$ = H, O, N, NH, OH and NH$_2$ (see panel B of Fig. \ref{fig:figure1}). 
An important point is that, after the attack of X$^\bullet$, the newly formed molecule is still a radical due to its unpaired electron and, therefore, reactive to couple with other open-shell species to possibly form iCOMs. In this work, the carbon bonded to the oxygen atom is labelled as C1 and the other as C2 (see Fig. \ref{fig:figure1}B).

All the studied reactions are exothermic, but present energy barriers (see Table \ref{tab:barriers}).

\begin{table*}[]
\centering
\caption{ZPE corrected reaction energies ($\Delta$H$_{react}$) and energy barriers ($\Delta$H$^\ddag$(0)) for every studied reaction, listed in kJ mol$^{-1}$ and computed with different levels of theory. 
$\omega$B97x, CCSD(T) and DLPNO stand for $\omega$B97x/TZVP, CCSD(T)-F12//$\omega$B97x/TZVP and DLPNO-F12//$\omega$B97x/TZVP level of theories, respectively.}
\label{tab:barriers}
\begin{tabular}{@{}lcccccc@{}}
\toprule
Reactions &
  \multicolumn{3}{c}{$\Delta$H$_{react}$} &
  \multicolumn{3}{c}{$\Delta$H$^\ddag$(0)} \\ \midrule
Attack on C1 &
  \multicolumn{1}{c}{$\omega$B97x} &
  \multicolumn{1}{c}{CCSD(T)} &
  \multicolumn{1}{c}{DLPNO} &
  \multicolumn{1}{c}{$\omega$B97x} &
  \multicolumn{1}{c}{CCSD(T)} &
  \multicolumn{1}{c}{DLPNO} \\ \midrule
Ketene + H & -147.0    & -147.6    & \multicolumn{1}{c|}{-150.1}  & 36.3    & 30.6          & 31.1   \\
Ketene + N & -75.4     & -46.4     & \multicolumn{1}{c|}{-49.5}    & 68.7    & 78.8          & 78.5   \\
Ketene + NH      & -116.5    & -99.9    & \multicolumn{1}{c|}{-101.3}   & 51.8    & 54.4          & 54.8   \\
Ketene + NH$_2$  & -169.1    & -163.8    & \multicolumn{1}{c|}{-163.1}   & 37.4    & 37.2          & 38.5   \\
Ketene + O       & -183.9    & -186.7    & \multicolumn{1}{c|}{-185.4}   & 14.9    & 15.7          & 17.1   \\
Ketene + OH      & -209.9    & -213.1    & \multicolumn{1}{c|}{-213.6}   & 5.2     & 3.3           & 4.6    \\ \midrule
Attack on C2 &  &  &  &  &  &  \\ \midrule
Ketene + H       & -173.9    & -175.6    & \multicolumn{1}{c|}{-175.7}   & 19.3    & 16.8          & 17.1   \\
Ketene + N       & -50.1     & -27.4     & \multicolumn{1}{c|}{-28.2}    & 62.7    & 74.8          & 74.1   \\
Ketene + NH      & -71.9     & -60.7     & \multicolumn{1}{c|}{-60.4}    & 59.0    & 63.9          & 63.9   \\
Ketene + NH$_2$  & -93.7     & -91.5      & \multicolumn{1}{c|}{-90.5}   & 43.9    & 43.5          & 43.4  \\
Ketene + O       & -121.3    & -115.3    & \multicolumn{1}{c|}{-115.7}   & 5.8     & 9.8           & 9.2    \\
Ketene + OH      & -117.5    & -123.1    & \multicolumn{1}{c|}{-122.9}   & 4.1     & 3.1           & 3.4    \\ \bottomrule
\end{tabular}
\end{table*}

In all cases, except for H, the products arising from the C1 attack are thermodynamically more stable than those from the C2 attack, due to the overstabilization gained by the $\pi$ delocalization when forming, for instance, an amide bond or a carboxylic group. 
However, C2 is the most prominent site to experience an attack due to the lower energy barriers in the cases of H, N, O and OH, whereas for NH and NH$_2$ the attack to C1 is preferred. Moreover, the OH radical attack presents the lowest energy barrier among the species studied, whereas N-bearing species are the most inert (see Table \ref{tab:barriers}). 

Note that DFT energy barriers are, in most cases, in good agreement with the results obtained with the CCSD(T)-F12 method. 
The barriers found for the hydrogenation reactions are also in good agreement with \cite{ibrahim_formation_2022}. 
Furthermore, it can be noticed that the DLPNO-F12 results agree fairly well with the CCSD(T)-F12 ones, which supports the use of DLPNO-F12 for the energetic refinement on CO ices, for which the CCSD(T)-F12 method cannot be employed due to the large size of the cluster.

\subsection{Solid phase calculations}
To confirm that the ketene formation is also doable in the solid state, the C($^3$P) + CO(ice) condensation was also investigated on the CO ice, resulting indeed in the formation of the \ch{C=C=O} ($^3\Sigma^-$) species.

Based on the gas-phase results, the cases of H (the most abundant species) and OH and NH$_2$ (the species presenting the lowest barriers for the O- and N-containing radical family species) have been selected to study their reactivity with ketene on the CO ice. 

\begin{figure*}
    \centering
    \includegraphics[width=\linewidth]{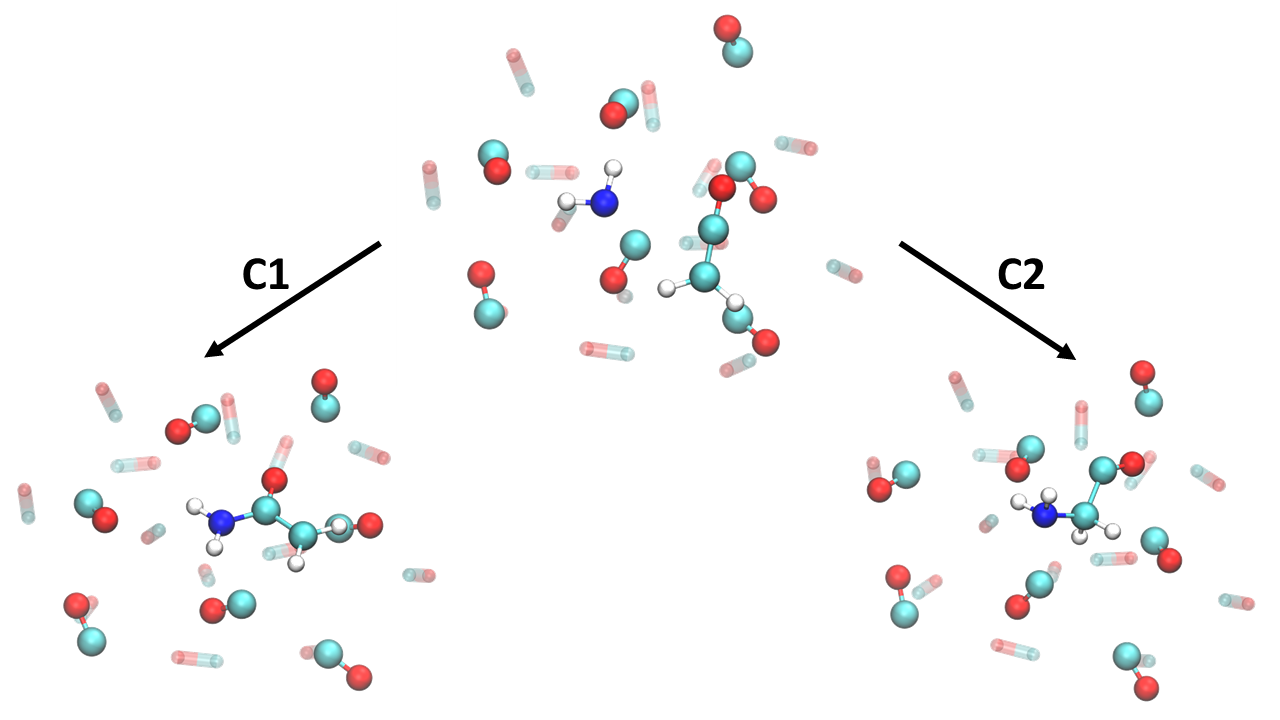}
    \caption{Optimized structures of the reactant and product for the ketene + NH$_2$ reaction taken as an example. Geometry relaxed species are represented in a ball and stick mode, whereas transparent CO molecules are held fixed during geometry optimization}
    \label{fig:reaction_onCO}
\end{figure*}

The reactions have been modeled by adopting a Langmuir-Hinshelwood mechanism. Thus, to calculate the reaction barriers, the reactant, TS and product structures have been optimized on two adjacent adsorption sites, including 7 unconstrained CO molecules (the rest remaining fixed, see Fig. \ref{fig:reaction_onCO}). The calculated barriers are reported in Table \ref{tab:CO_solid_barriers}.
According to the ISM conditions, they are all very high (the lowest one being 9.3 kJ mol$^{-1}$) for the reactions to proceed.
In the following, we discuss the differences with the barriers obtained in the gas phase.

\begin{table}[ht]
\centering
\caption{Computed ZPE-corrected energy barriers ($\Delta$H$^\ddag$(0)) for the reactions of ketene with H, NH$_2$ and OH on the CO cluster considering the two C atoms of ketene (C1 and C2). Values are reported in kJ mol$^{-1}$ }
\label{tab:CO_solid_barriers}
\begin{tabular}{lcc}
\toprule
\multicolumn{1}{c}{Species} & Attack on C1 $\Delta$H$^\ddag$(0) & Attack on C2 $\Delta$H$^\ddag$(0)\\ \midrule
                            & DLPNO-F12    & DLPNO-F12    \\ \midrule
H                           & 29.1         & 13.5         \\
NH$_2$                         & 41.0         & 42.9         \\
OH                          & 10.2         & 9.3          \\ \bottomrule
\end{tabular}
\end{table}

\textit{Ketene + NH$_2$:} Comparison of the solid-state barriers with the gas-phase ones gives differences that are less than 4 kJ mol$^{-1}$. Since the energy barriers for both cases are high (insurmountable in the ISM conditions), these variations have no practical effects so that CO ice behaves as an inert surface. 

\textit{Ketene + OH:} For OH, there is an increase of 6 kJ mol$^{-1}$ on CO, a non-negligible variation since the gas-phase barriers are very low. 
The increase can be attributed to the interactions between OH and the CO ice (absent in the gas phase), which need to be partly overcome to proceed with the reaction on the surface. Interestingly, for this case, the energy barrier is relatively small, which could be overcome classically by means of non-thermal mechanisms (as reported in works involving reactive OH radicals \citep[see][]{Pauly2011, Ishibashi2021}), or could even proceed with the help of heavy atom quantum tunneling below the crossover temperature \citep[see][]{Castro, MeisnerKastner_angewandte}. Moreover, it is worth mentioning that the calculated barrier arises from just one starting configuration of the reactants, which could well be lower in other surface reactive sites.

\textit{Ketene + H:} Here, the C2 attack appears to be the most doable as it presents a barrier that is less than half of the barrier of the C1 reaction but it is still very high for the ISM conditions. However, it is worth noticing that reactions happening at low temperatures and involving light species such as H, tunneling effects can dominate and, therefore, considering only classical barrier heights can lead to wrong conclusions. This kinetic aspect is discussed in more detail in the next section.

Another crucial point for the on-grain reactivity is the species/ice interactions that can be established at the surfaces. 
On icy water grains, species like NH$_2$ and OH can form strong hydrogen bond (H-bond) interactions, inhibiting their diffusivity and, hence, their reactivity via Langmuir-Hinshelwood. These interactions have been studied on water ice surfaces both experimentally and theoretically \citep{sameera_modelling_2022, sameera_oniomqmamoeba09_2017, tsuge_behavior_2021, enrique-romero_reactivity_2019, enrique-romero_quantum_2022, enrique-romero_theoretical_2021, ferrero_binding_2020, wakelam_binding_2017, duflot_theoretical_2021}.
However, very little data are available in the literature relative to CO ices. 
To have an idea of the interactions of OH and NH$_2$ with CO-rich ices (different in nature to water ice), we used a similar strategy as in \citet{lamberts_formation_2019} to calculate interaction energies: we sampled ten different binding sites around the CO cluster to calculate the corresponding $\Delta$H(0) adsorption energies (results shown in Table \ref{tab:BEs_CO}).

\begin{table}
    \caption{Adsorption energies on CO- and H$_2$O ices of NH$_2$ and OH:
    mean value and standard deviation of the adsorption energies $\Delta$H(0) on the CO cluster calculated in this work are reported (second column) and calculated data on different water ice models from other works (third column). 
    When more than two values were present in the literature, mean and standard deviation have been calculated. 
    All values are in kJ mol$^{-1}$.}
\label{tab:BEs_CO}
\begin{tabular}{ccc}
\toprule
Species & CO ice & Water ice \\ 
\midrule
OH      & -22.1$\pm$ 3.9 & -43.2$\pm$16.1$^a$;\\
        &                & -44.4$^b$\\
        &                & -32 -- -41$^c$\\
        &                & -26.9$\pm$11.5$^d$\\
NH$_2$  & -6.5$\pm$ 3.2  & -31 -- -44$^e$\\
        &                &-27.9$\pm$3.1$^d$ \\ 
\bottomrule
\end{tabular}
\tablecomments{References: $^a$, \cite{duflot_theoretical_2021}; $^b$, \cite{sameera_oniomqmamoeba09_2017}; $^c$, \cite{meisner_atom_2017}; $^d$, \cite{ferrero_binding_2020}; $^e$, \cite{enrique-romero_reactivity_2019}.}
\end{table}

Although the binding site sampling is not exhaustive, we can compare our values with other computational works that calculated the interaction of these two species on water ice models. 
The difference in the adsorption energies is quite large, clearly indicating that the interaction on CO-rich ices is significantly weaker than on H$_2$O-rich ones (particularly for NH$_2$), as already found in previous works for different radicals \citep{lamberts_formation_2019}. 
This is due to the fewer and weaker H-bonds that OH and NH$_2$ form on CO ice with respect to water ice. This fact is also connected with the mobility of these species on different icy surfaces. Even without calculating explicitly the diffusion energy barriers, it is noticeable that the diffusivity of these two radicals on CO-rich ices will be larger than that on water ices due to the weaker radical/surface interactions in the former. However, the binding energies of these species on CO ice are still significantly large so diffusion is expected to be very slow at the ISM conditions, which in turn hamper diffusion-limited reactions like radical-radical couplings.



\subsection{Kinetic calculations and tunneling effects}
\begin{figure*}[ht]
    \centering
    \includegraphics[width=\linewidth]{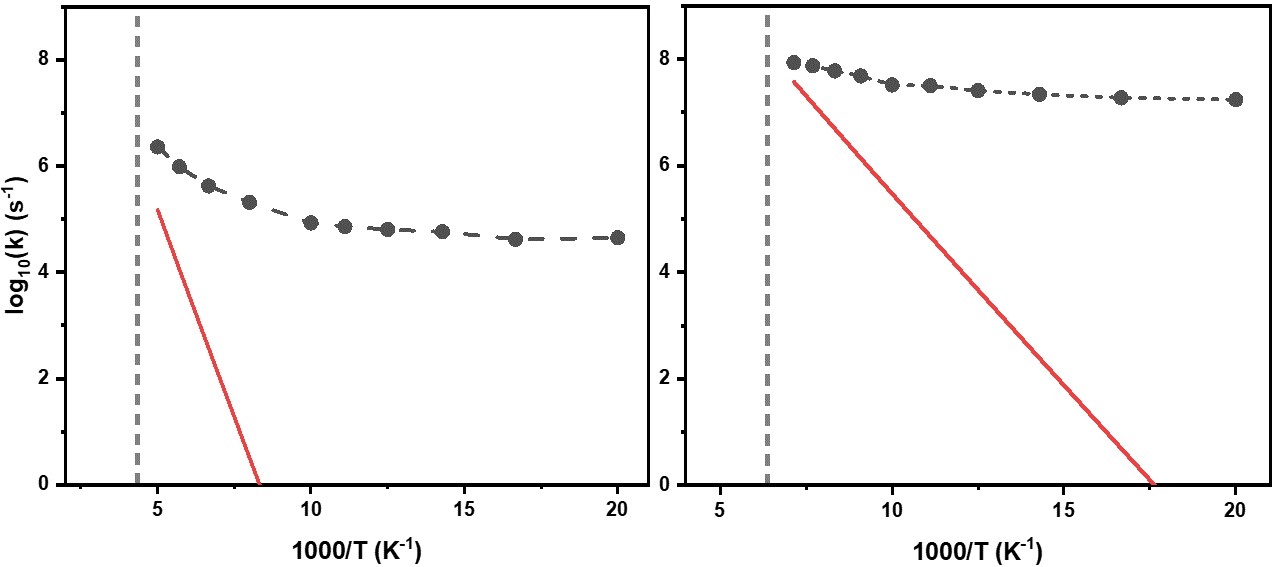}
    \caption{Arrhenius plots for the two different hydrogenation reactions, on C1 (left panel) and C2 (right panel). Black dots correspond to values of the rate constant calculated with instanton theory at different temperatures whereas the classical Eyring rate constant is depicted in red. Gray dashed vertical lines denote the crossover temperature.}
    \label{fig:instanton}
\end{figure*}

At the cryogenic temperatures of the ISM and for reactions involving light atoms, like H, quantum tunneling cannot be overlooked. 
Thus, ketene hydrogenation considering both C1 and C2 attacks was also studied adopting the instanton theory (outlined in section \ref{sec:inst_rate_theory}), which calculates the most probable tunneling path connecting reactants and products. 
Regarding the attack on C1, the calculated T$_c$ is 230 K, whereas on C2 is 157K. Figure \ref{fig:instanton} reports the Arrhenius plots for these two H additions. 
As done in previous works \citep{meisner_atom_2017, lamberts_influence_2017, lamberts_quantum_2016}, the reactions are considered as unimolecular because the diffusion of the two reactants is not considered and the rate constants measure the rate starting from a reactive complex. 
The calculated rate constants refined at CCSD(T)-F12/cc-pVTZ-F12//$\omega$B97X-D4/TZVP at 50 K are $4.47 \times 10^4$ s$^{-1}$ and $1.74 \times 10^7$ s$^{-1}$ for C1 and C2 hydrogenations, respectively. Remarkably, rate constants keep almost invariant at these temperatures, so we can assume similar values for 10 K. Therefore, both attacks are viable at ISM temperatures, being on C2 kinetically more favorable. 
As stated in section \ref{sec:inst_rate_theory}, these instanton calculations have been carried out in the gas phase, but the surface effects have been considered through the implicit surface approximation, which in this case holds because the CO surface acts as an inert substrate without affecting significantly the classical barriers.

\section{Astrophysical implications and concluding remarks}

The calculations presented here show that, when atomic C lands on an icy CO surface, \ch{C=C=O} forms, as in the gas phase \citep{papakondylis_electronic_2019}.
However, once formed, it could very likely be hydrogenated into ketene by the addition of H, also landing on the grain surfaces.
We then explored if ketene (CH$_2$CO) could grow into larger molecules by reacting with simple and abundant radicals on the grain surfaces, namely H, OH and NH$_2$.

These reactions on the (non-polar) CO ice present relatively high energy barriers ($\geq 9$ kJ/mol) for the ISM conditions. It seems clear that ketene hydrogenation can occur through  the H tunneling. Thus, acetyl radical (CH$_3$CO) formation (through H addition to ketene) is the most kinetically favourable path. Formation of radical precursors for acetamide (CH$_3$CONH$_2$) and acetic acid (CH$_3$COOH) (through NH$_2$ and OH addition, respectively), according to our results, are not expected to form. However, local surface heating and other non-thermal mechanisms could be operative, this way helping the occurrence of these reactions, so that these synthetic paths cannot be discarded.

CH$_3$CO can be in turn hydrogenated and produce acetaldehyde (CH$_3$CHO) and/or CO + CH$_4$.
Recent experiments indicate that the latter channel competes with the first one, with a branching ratio up to 80\%, against naive expectations \citep{ibrahim_formation_2022}.
One can speculate that acetaldehyde can successively be hydrogenated into ethanol by other H additions. 
One could also suppose that the CH$_3$CO radical could react with the OH and NH$_2$ radicals. However, our computed relatively large binding energies of OH and NH$_2$ on CO ice point towards a very low diffusivity, inhibiting, once again, the formation of CH$_3$COOH and CH$_3$CONH$_2$ via radical-radical coupling with CH$_3$CO.

As mentioned in the Introduction, gaseous atomic C and frozen CO are simultaneously found in molecular clouds, probably in the photo-dissociated region (PDR) skin. 
The bottleneck to form ketene is probably the quantity of gaseous atomic C, since a substantial fraction of frozen CO, about 25\% of the ice, is observed in diffuse clouds \citep[with visual extinction $\sim3$ mag:][]{boogert_observations_2015}.
Estimating the amount of gaseous atomic C, where also frozen CO exists, is not easy, but observations show that it is up to the same abundance of gaseous CO in giant molecular clouds \citep[e.g.][]{Plume1999}, making the \ch{C=C=O} hydrogenation a possible important source of frozen acetaldehyde and, perhaps, ethanol.
It is worth emphasizing that in the interiors of the molecular clouds, where carbon is almost entirely trapped in CO, neutral carbon is not expected to be as abundant as in their skins. 
Therefore, the mechanism described in this work is mostly viable in the latter, which represents only a small fraction of the molecular clouds.
That said, little is known of the composition of ices in the PDRs and whether surface chemistry, other than the formation of H$_2$, is efficient.
Indirect proof that frozen CO is present and that it is hydrogenated is provided by the observed relatively abundant gaseous methanol in PDRs \citep[e.g.][]{Bouvier2020-PDRs}.
In this respect, therefore, the \ch{C=C=O} hydrogenation probably occurs in these regions and can lead to frozen acetaldehyde and, perhaps, ethanol.

Another possibility is that ketene formed in the gas phase is also frozen into the grain icy mantles.
Astrochemical models predict a gaseous ketene abundance of about $10^{-10}$--$10^{-8}$ while observations indicate  $10^{-10}$--$10^{-9}$ abundances \citep[e.g.,][]{bacmann2012,jaber2014,vastel2014}.
Assuming that the frozen ketene is all converted into acetaldehyde and/or ethanol, it provides upper limits to their abundance of $10^{-10}$--$10^{-9}$ with respect to H, namely about $10^{-6}$--$10^{-5}$ with respect to frozen water.

In summary, it is possible that (frozen) ketene is formed in the PDR skins of the molecular clouds, where C and frozen CO may coexist, which then would trigger the formation of iced acetaldehyde and ethanol, possibly explaining the new observations of JWST (which need confirmation).

\section{acknowledgments}
S.F. is deeply indebted to Gabriel Laude and Jeremy Richardson for their help and advice on instanton calculations. This project has received funding from the Marie Sklodowska-Curie for the project ``Astro-Chemical Origins” (ACO), grant agreement No 811312 and within the European Union’s Horizon 2020 research and innovation program from the European Research Council (ERC) for the projects ``The Dawn of Organic Chemistry” (DOC), grant agreement No 741002 and ``Quantum Chemistry on Interstellar Grains” (QUANTUMGRAIN), grant agreement No 865657.




\bibliography{bibliography}{}

\begin{thebibliography}{}
\expandafter\ifx\csname natexlab\endcsname\relax\def\natexlab#1{#1}\fi
\providecommand{\url}[1]{\href{#1}{#1}}

\bibitem[{Adler {et~al.}(2007)Adler, Knizia, \& Werner}]{adler_simple_2007}
Adler, T.~B., Knizia, G., \& Werner, H.-J. 2007, \jcp, 127, 221106.
\newblock \url{https://aip.scitation.org/doi/full/10.1063/1.2817618}

\bibitem[{Andersson {et~al.}(2009)Andersson, Nyman, Arnaldsson, Manthe, \&
  J{\'o}nsson}]{andersson_comparison_2009}
Andersson, S., Nyman, G., Arnaldsson, A., Manthe, U., \& J{\'o}nsson, H. 2009,
  J. Phys. Chem. A, 113, 4468.
\newblock \url{https://doi.org/10.1021/jp811070w}

\bibitem[{{Bacmann} {et~al.}(2002){Bacmann}, {Lefloch}, {Ceccarelli},
  {Castets}, {Steinacker}, \& {Loinard}}]{bacmann_degree_2002}
{Bacmann}, A., {Lefloch}, B., {Ceccarelli}, C., {et~al.} 2002, \aap, 389, L6.
\newblock \url{https://ui.adsabs.harvard.edu/abs/2002A&A...389L...6B}

\bibitem[{{Bacmann} {et~al.}(2012){Bacmann}, {Taquet}, {Faure}, {Kahane}, \&
  {Ceccarelli}}]{bacmann2012}
{Bacmann}, A., {Taquet}, V., {Faure}, A., {Kahane}, C., \& {Ceccarelli}, C.
  2012, \aap, 541, L12.
\newblock \url{https://ui.adsabs.harvard.edu/abs/2012A&A...541L..12B}

\bibitem[{Beyer {et~al.}(2016)Beyer, Richardson, Knowles, Rommel, \&
  Althorpe}]{beyer_quantum_2016}
Beyer, A.~N., Richardson, J.~O., Knowles, P.~J., Rommel, J., \& Althorpe, S.~C.
  2016, J Phys Chem Lett, 7, 4374.
\newblock \url{https://doi.org/10.1021/acs.jpclett.6b02115}

\bibitem[{Boogert {et~al.}(2015)Boogert, Gerakines, \&
  Whittet}]{boogert_observations_2015}
Boogert, A.~A., Gerakines, P.~A., \& Whittet, D.~C. 2015, \araa, 53, 541.
\newblock \url{https://doi.org/10.1146/annurev-astro-082214-122348}

\bibitem[{{Bouvier} {et~al.}(2020){Bouvier}, {L{\'o}pez-Sepulcre},
  {Ceccarelli}, {Kahane}, {Imai}, {Sakai}, {Yamamoto}, \&
  {Dagdigian}}]{Bouvier2020-PDRs}
{Bouvier}, M., {L{\'o}pez-Sepulcre}, A., {Ceccarelli}, C., {et~al.} 2020, \aap,
  636, A19

\bibitem[{Caselli {et~al.}(1999)Caselli, Walmsley, Tafalla, Dore, \&
  Myers}]{caselli_co_1999}
Caselli, P., Walmsley, C.~M., Tafalla, M., Dore, L., \& Myers, P.~C. 1999, ApJ,
  523, L165.
\newblock \url{https://ui.adsabs.harvard.edu/abs/1999ApJ...523L.165C}

\bibitem[{Castro \& Karney(2020)}]{Castro}
Castro, C., \& Karney, W.~L. 2020, Angewandte Chemie International Edition, 59,
  8355.
\newblock \url{https://onlinelibrary.wiley.com/doi/abs/10.1002/anie.201914943}

\bibitem[{Ceccarelli {et~al.}(2022)Ceccarelli, Codella, Balucani,
  BockelÃ©e-Morvan, Herbst, Vastel, Caselli, Favre, Lefloch, \&
  Ã–berg}]{ceccarelli_organic_2022}
Ceccarelli, C., Codella, C., Balucani, N., {et~al.} 2022, Organic chemistry in
  the first phases of {Solar}-type protostars, Tech. rep.
\newblock \url{https://ui.adsabs.harvard.edu/abs/2022arXiv220613270C}

\bibitem[{Chapman {et~al.}(1975)Chapman, Garrett, \&
  Miller}]{chapman_semiclassical_1975}
Chapman, S., Garrett, B.~C., \& Miller, W.~H. 1975, \jcp, 63, 2710.
\newblock \url{https://aip.scitation.org/doi/10.1063/1.431620}

\bibitem[{Chuang {et~al.}(2017)Chuang, Fedoseev, Qasim, Ioppolo, van Dishoeck,
  \& Linnartz}]{chuang_production_2017}
Chuang, K.~J., Fedoseev, G., Qasim, D., {et~al.} 2017, \mnras, 467, 2552.
\newblock \url{https://ui.adsabs.harvard.edu/abs/2017MNRAS.467.2552C}

\bibitem[{{Chuang} {et~al.}(2021){Chuang}, {Fedoseev}, {Scir{\`e}}, {Baratta},
  {J{\"a}ger}, {Henning}, {Linnartz}, \& {Palumbo}}]{chuang_formation_2021}
{Chuang}, K.~J., {Fedoseev}, G., {Scir{\`e}}, C., {et~al.} 2021, \aap, 650,
  A85.
\newblock \url{https://ui.adsabs.harvard.edu/abs/2021A&A...650A..85C}

\bibitem[{Cuppen {et~al.}(2017)Cuppen, Walsh, Lamberts, Semenov, Garrod,
  Penteado, \& Ioppolo}]{cuppen_grain_2017}
Cuppen, H.~M., Walsh, C., Lamberts, T., {et~al.} 2017, SSRv, 212, 1.
\newblock \url{https://doi.org/10.1007/s11214-016-0319-3}

\bibitem[{Duflot {et~al.}(2021)Duflot, Toubin, \&
  Monnerville}]{duflot_theoretical_2021}
Duflot, D., Toubin, C., \& Monnerville, M. 2021, Front. Astron. Space Sci., 8.
\newblock \url{https://www.frontiersin.org/articles/10.3389/fspas.2021.645243}

\bibitem[{Enrique-Romero {et~al.}(2021)Enrique-Romero, Ceccarelli, Rimola,
  Skouteris, Balucani, \& Ugliengo}]{enrique-romero_theoretical_2021}
Enrique-Romero, J., Ceccarelli, C., Rimola, A., {et~al.} 2021, \aap, 655, A9.
\newblock
  \url{https://www.aanda.org/articles/aa/abs/2021/11/aa41531-21/aa41531-21.html}

\bibitem[{Enrique-Romero {et~al.}(2019)Enrique-Romero, Rimola, Ceccarelli,
  Ugliengo, Balucani, \& Skouteris}]{enrique-romero_reactivity_2019}
Enrique-Romero, J., Rimola, A., Ceccarelli, C., {et~al.} 2019, ACS Earth Space
  Chem., 3, 2158.
\newblock \url{https://doi.org/10.1021/acsearthspacechem.9b00156}

\bibitem[{Enrique-Romero {et~al.}(2022)Enrique-Romero, Rimola, Ceccarelli,
  Ugliengo, Balucani, \& Skouteris}]{enrique-romero_quantum_2022}
---. 2022, ApJS, 259, 39.
\newblock \url{https://doi.org/10.3847/1538-4365/ac480e}

\bibitem[{Favre {et~al.}(2013)Favre, Cleeves, Bergin, Qi, \&
  Blake}]{favre_significantly_2013}
Favre, C., Cleeves, L.~I., Bergin, E.~A., Qi, C., \& Blake, G.~A. 2013, ApJ,
  776, L38.
\newblock \url{https://ui.adsabs.harvard.edu/abs/2013ApJ...776L..38F}

\bibitem[{Fedoseev {et~al.}(2022)Fedoseev, Qasim, Chuang, Ioppolo, Lamberts,
  Dishoeck, \& Linnartz}]{fedoseev_hydrogenation_2022}
Fedoseev, G., Qasim, D., Chuang, K.-J., {et~al.} 2022, ApJ, 924, 110.
\newblock \url{https://doi.org/10.3847/1538-4357/ac3834}

\bibitem[{Ferrero {et~al.}(2020)Ferrero, Zamirri, Ceccarelli, Witzel, Rimola,
  \& Ugliengo}]{ferrero_binding_2020}
Ferrero, S., Zamirri, L., Ceccarelli, C., {et~al.} 2020, ApJ, 904, 11.
\newblock
  \url{https://iopscience.iop.org/article/10.3847/1538-4357/abb953/meta}

\bibitem[{{Garrod} \& {Herbst}(2006)}]{Garrod2006}
{Garrod}, R.~T., \& {Herbst}, E. 2006, \aap, 457, 927.
\newblock \url{https://ui.adsabs.harvard.edu/abs/2006A&A...457..927G}

\bibitem[{{Garrod} \& {Pauly}(2011)}]{Pauly2011}
{Garrod}, R.~T., \& {Pauly}, T. 2011, \apj, 735, 15

\bibitem[{H{\"a}ttig {et~al.}(2012)H{\"a}ttig, Klopper, K{\"o}hn, \&
  Tew}]{hattig_explicitly_2012}
H{\"a}ttig, C., Klopper, W., K{\"o}hn, A., \& Tew, D.~P. 2012, Chem Rev, 112,
  4.
\newblock \url{https://doi.org/10.1021/cr200168z}

\bibitem[{Henning \& Krasnokutski(2019)}]{henning_experimental_2019}
Henning, T.~K., \& Krasnokutski, S.~A. 2019, Nat. Astron., 3, 568.
\newblock \url{https://www.nature.com/articles/s41550-019-0729-8}

\bibitem[{Humphrey {et~al.}(1996)Humphrey, Dalke, \&
  Schulten}]{humphrey_vmd_1996}
Humphrey, W., Dalke, A., \& Schulten, K. 1996, J. Mol. Graph., 14, 33.
\newblock
  \url{https://www.sciencedirect.com/science/article/pii/0263785596000185}

\bibitem[{Ibrahim {et~al.}(2022)Ibrahim, Guillemin, Chaquin, Markovits, \&
  Krim}]{ibrahim_formation_2022}
Ibrahim, M., Guillemin, J.-C., Chaquin, P., Markovits, A., \& Krim, L. 2022,
  PCCP, doi:10.1039/D2CP02980D.
\newblock
  \url{https://pubs.rsc.org/en/content/articlelanding/2022/cp/d2cp02980d}

\bibitem[{{Ishibashi} {et~al.}(2021){Ishibashi}, {Hidaka}, {Oba}, {Kouchi}, \&
  {Watanabe}}]{Ishibashi2021}
{Ishibashi}, A., {Hidaka}, H., {Oba}, Y., {Kouchi}, A., \& {Watanabe}, N. 2021,
  \apjl, 921, L13

\bibitem[{{Jaber} {et~al.}(2014){Jaber}, {Ceccarelli}, {Kahane}, \&
  {Caux}}]{jaber2014}
{Jaber}, A.~A., {Ceccarelli}, C., {Kahane}, C., \& {Caux}, E. 2014, \apj, 791,
  29.
\newblock \url{https://ui.adsabs.harvard.edu/abs/2014ApJ...791...29J}

\bibitem[{{Kamegai} {et~al.}(2003){Kamegai}, {Ikeda}, {Maezawa}, {Ito},
  {Iwata}, {Sakai}, {Oka}, {Yamamoto}, {Sekimoto}, {Tatematsu}, {Noguchi},
  {Saito}, {Fujiwara}, {Ozeki}, {Inatani}, \& {Ohishi}}]{Kamegai2003}
{Kamegai}, K., {Ikeda}, M., {Maezawa}, H., {et~al.} 2003, \apj, 589, 378.
\newblock \url{https://ui.adsabs.harvard.edu/abs/2003ApJ...589..378K}

\bibitem[{K{\"a}stner(2014)}]{kastner_theory_2014}
K{\"a}stner, J. 2014, Wiley Interdiscip. Rev. Comput. Mol. Sci., 4, 158.
\newblock \url{https://onlinelibrary.wiley.com/doi/abs/10.1002/wcms.1165}

\bibitem[{Kong {et~al.}(2012)Kong, Bischoff, \& Valeev}]{kong_explicitly_2012}
Kong, L., Bischoff, F.~A., \& Valeev, E.~F. 2012, Chem Rev, 112, 75.
\newblock \url{https://doi.org/10.1021/cr200204r}

\bibitem[{Krasnokutski {et~al.}(2022)Krasnokutski, Chuang, J{\"a}ger,
  Ueberschaar, \& Henning}]{krasnokutski_pathway_2022}
Krasnokutski, S.~A., Chuang, K.-J., J{\"a}ger, C., Ueberschaar, N., \& Henning,
  T. 2022, Nat. Astron., 1.
\newblock \url{https://www.nature.com/articles/s41550-021-01577-9}

\bibitem[{Krasnokutski {et~al.}(2020)Krasnokutski, J{\"a}ger, \&
  Henning}]{krasnokutski_condensation_2020}
Krasnokutski, S.~A., J{\"a}ger, C., \& Henning, T. 2020, ApJ, 889, 67.
\newblock \url{https://doi.org/10.3847/1538-4357/ab60a1}

\bibitem[{Krasnokutski {et~al.}(2017)Krasnokutski, Goulart, Gordon, Ritsch,
  J{\"a}ger, Rastogi, Salvenmoser, Henning, \&
  Scheier}]{krasnokutski_low-temperature_2017}
Krasnokutski, S.~A., Goulart, M., Gordon, E.~B., {et~al.} 2017, ApJ, 847, 89.
\newblock \url{https://ui.adsabs.harvard.edu/abs/2017ApJ...847...89K}

\bibitem[{Kruse \& Grimme(2012)}]{kruse_geometrical_2012}
Kruse, H., \& Grimme, S. 2012, \jcp, 136, 154101.
\newblock \url{https://aip.scitation.org/doi/10.1063/1.3700154}

\bibitem[{Lamberts \& K{\"a}stner(2017)}]{lamberts_influence_2017}
Lamberts, T., \& K{\"a}stner, J. 2017, ApJ, 846, 43.
\newblock \url{https://doi.org/10.3847/1538-4357/aa8311}

\bibitem[{Lamberts {et~al.}(2016)Lamberts, Kumar Samanta, K{\"o}hn, \&
  K{\"a}stner}]{lamberts_quantum_2016}
Lamberts, T., Kumar Samanta, P., K{\"o}hn, A., \& K{\"a}stner, J. 2016, PCCP,
  18, 33021.
\newblock
  \url{https://pubs.rsc.org/en/content/articlelanding/2016/cp/c6cp06457d}

\bibitem[{Lamberts {et~al.}(2019)Lamberts, Markmeyer, Kolb, \&
  K{\"a}stner}]{lamberts_formation_2019}
Lamberts, T., Markmeyer, M.~N., Kolb, F.~J., \& K{\"a}stner, J. 2019, ACS Earth
  Space Chem., 3, 958.
\newblock \url{https://doi.org/10.1021/acsearthspacechem.9b00029}

\bibitem[{Li \& Jensen(2002)}]{li2002partial}
Li, H., \& Jensen, J.~H. 2002, Theoretical Chemistry Accounts, 107, 211

\bibitem[{Liu \& McLean(1973)}]{liu_accurate_1973}
Liu, B., \& McLean, A.~D. 1973, \jcp, 59, 4557.
\newblock \url{https://aip.scitation.org/doi/10.1063/1.1680654}

\bibitem[{{McClure} {et~al.}(2023){McClure}, {Rocha}, {Pontoppidan}, {Crouzet},
  {Chu}, {Dartois}, {Lamberts}, {Noble}, {Pendleton}, {Perotti}, {Qasim},
  {Rachid}, {Smith}, {Sun}, {Beck}, {Boogert}, {Brown}, {Caselli}, {Charnley},
  {Cuppen}, {Dickinson}, {Drozdovskaya}, {Egami}, {Erkal}, {Fraser}, {Garrod},
  {Harsono}, {Ioppolo}, {Jim{\'e}nez-Serra}, {Jin}, {J{\o}rgensen},
  {Kristensen}, {Lis}, {McCoustra}, {McGuire}, {Melnick}, {{\"O}berg},
  {Palumbo}, {Shimonishi}, {Sturm}, {van Dishoeck}, \&
  {Linnartz}}]{McClure2023NatAs}
{McClure}, M.~K., {Rocha}, W.~R.~M., {Pontoppidan}, K.~M., {et~al.} 2023, Nat.
  Astron., arXiv:2301.09140.
\newblock \url{https://ui.adsabs.harvard.edu/abs/2023NatAs.tmp...25M}

\bibitem[{Meisner \& K{\"a}stner(2018)}]{meisner_dual-level_2018}
Meisner, J., \& K{\"a}stner, J. 2018, JCTC, 14, 1865.
\newblock \url{https://doi.org/10.1021/acs.jctc.8b00068}

\bibitem[{Meisner \& Kästner(2016)}]{MeisnerKastner_angewandte}
Meisner, J., \& Kästner, J. 2016, Angewandte Chemie International Edition, 55,
  5400.
\newblock \url{https://onlinelibrary.wiley.com/doi/abs/10.1002/anie.201511028}

\bibitem[{Meisner {et~al.}(2017)Meisner, Lamberts, \&
  K{\"a}stner}]{meisner_atom_2017}
Meisner, J., Lamberts, T., \& K{\"a}stner, J. 2017, ACS Earth Space Chem., 1,
  399.
\newblock \url{https://doi.org/10.1021/acsearthspacechem.7b00052}

\bibitem[{Miller(1975)}]{miller_semiclassical_1975}
Miller, W.~H. 1975, \jcp, 62, 1899.
\newblock \url{https://aip.scitation.org/doi/10.1063/1.430676}

\bibitem[{Molpeceres {et~al.}(2021)Molpeceres, K{\"a}stner, Fedoseev, Qasim,
  SchÃ¶mig, Linnartz, \& Lamberts}]{molpeceres_carbon_2021}
Molpeceres, G., K{\"a}stner, J., Fedoseev, G., {et~al.} 2021, J. Phys. Chem.
  Lett., 12, 10854.
\newblock \url{https://doi.org/10.1021/acs.jpclett.1c02760}

\bibitem[{Najibi \& Goerigk(2020)}]{najibi_dft-d4_2020}
Najibi, A., \& Goerigk, L. 2020, J. Comput. Chem., 41, 2562.
\newblock \url{https://onlinelibrary.wiley.com/doi/abs/10.1002/jcc.26411}

\bibitem[{Neese {et~al.}(2020)Neese, Wennmohs, Becker, \&
  Riplinger}]{neese_orca_2020}
Neese, F., Wennmohs, F., Becker, U., \& Riplinger, C. 2020, \jcp, 152, 224108.
\newblock \url{https://aip.scitation.org/doi/10.1063/5.0004608}

\bibitem[{Papakondylis \& Mavridis(2019)}]{papakondylis_electronic_2019}
Papakondylis, A., \& Mavridis, A. 2019, J. Phys. Chem. A, 123, 10290.
\newblock \url{https://pubs.acs.org/doi/10.1021/acs.jpca.9b09084}

\bibitem[{Pavo{\v s}evi{\'c} {et~al.}(2017)Pavo{\v s}evi{\'c}, Peng, Pinski,
  Riplinger, Neese, \& Valeev}]{pavosevic_sparsemapssystematic_2017}
Pavo{\v s}evi{\'c}, F., Peng, C., Pinski, P., {et~al.} 2017, \jcp, 146, 174108.
\newblock \url{https://aip.scitation.org/doi/full/10.1063/1.4979993}

\bibitem[{Peterson {et~al.}(2008)Peterson, Adler, \&
  Werner}]{peterson_systematically_2008}
Peterson, K.~A., Adler, T.~B., \& Werner, H.-J. 2008, \jcp, 128, 084102.
\newblock \url{https://aip.scitation.org/doi/abs/10.1063/1.2831537}

\bibitem[{{Plume} {et~al.}(1999){Plume}, {Jaffe}, {Tatematsu}, {Evans}, \&
  {Keene}}]{Plume1999}
{Plume}, R., {Jaffe}, D.~T., {Tatematsu}, K., {Evans}, Neal~J., I., \& {Keene},
  J. 1999, \apj, 512, 768.
\newblock \url{https://ui.adsabs.harvard.edu/abs/1999ApJ...512..768P}

\bibitem[{Pontoppidan {et~al.}(2008)Pontoppidan, Boogert, Fraser, van Dishoeck,
  Blake, Lahuis, Ã–berg, Evans, \& Salyk}]{pontoppidan_c2d_2008}
Pontoppidan, K.~M., Boogert, A. C.~A., Fraser, H.~J., {et~al.} 2008, ApJ, 678,
  1005.
\newblock \url{https://ui.adsabs.harvard.edu/abs/2008ApJ...678.1005P}

\bibitem[{Potapov {et~al.}(2021)Potapov, Krasnokutski, J{\"a}ger, \&
  Henning}]{potapov_new_2021}
Potapov, A., Krasnokutski, S.~A., J{\"a}ger, C., \& Henning, T. 2021, ApJ, 920,
  111.
\newblock \url{https://doi.org/10.3847/1538-4357/ac1a70}

\bibitem[{Potapov \& McCoustra(2021)}]{McCoustra-Rev}
Potapov, A., \& McCoustra, M. 2021, Int Rev Phys Chem, 40, 299.
\newblock \url{https://doi.org/10.1080/0144235X.2021.1918498}

\bibitem[{Richardson(2018{\natexlab{a}})}]{richardson_perspective_2018}
Richardson, J.~O. 2018{\natexlab{a}}, \jcp, 148, 200901.
\newblock \url{https://aip.scitation.org/doi/10.1063/1.5028352}

\bibitem[{Richardson(2018{\natexlab{b}})}]{richardson_ring-polymer_2018}
---. 2018{\natexlab{b}}, Int Rev Phys Chem, 37, 171.
\newblock \url{https://doi.org/10.1080/0144235X.2018.1472353}

\bibitem[{Sameera {et~al.}(2017)Sameera, Senevirathne, Andersson, Maseras, \&
  Nyman}]{sameera_oniomqmamoeba09_2017}
Sameera, W. M.~C., Senevirathne, B., Andersson, S., Maseras, F., \& Nyman, G.
  2017, J Phys Chem C, 121, 15223.
\newblock \url{https://doi.org/10.1021/acs.jpcc.7b04105}

\bibitem[{Sameera {et~al.}(2022)Sameera, Senevirathne, Nguyen, Oba, Ishibashi,
  Tsuge, Hidaka, \& Watanabe}]{sameera_modelling_2022}
Sameera, W. M.~C., Senevirathne, B., Nguyen, T., {et~al.} 2022, Front. Astron.
  Space Sci., 9.
\newblock \url{https://www.frontiersin.org/articles/10.3389/fspas.2022.890161}

\bibitem[{Shimonishi {et~al.}(2018)Shimonishi, Nakatani, Furuya, \&
  Hama}]{shimonishi_adsorption_2018}
Shimonishi, T., Nakatani, N., Furuya, K., \& Hama, T. 2018, ApJ, 855, 27.
\newblock \url{https://doi.org/10.3847/1538-4357/aaaa6a}

\bibitem[{Simons {et~al.}(2020)Simons, Lamberts, \&
  Cuppen}]{simons_formation_2020}
Simons, M. a.~J., Lamberts, T., \& Cuppen, H.~M. 2020, \aap, 634, A52.
\newblock \url{https://ui.adsabs.harvard.edu/abs/2020A&A...634A..52S/abstract}

\bibitem[{{Skouteris} {et~al.}(2018){Skouteris}, {Balucani}, {Ceccarelli},
  {Vazart}, {Puzzarini}, {Barone}, {Codella}, \&
  {Lefloch}}]{Skouteris2018-ethanoltree}
{Skouteris}, D., {Balucani}, N., {Ceccarelli}, C., {et~al.} 2018, \apj, 854,
  135.
\newblock \url{https://ui.adsabs.harvard.edu/abs/2018ApJ...854..135S}

\bibitem[{{Tielens} \& {Hagen}(1982)}]{Tielens1982}
{Tielens}, A.~G.~G.~M., \& {Hagen}, W. 1982, \aap, 114, 245.
\newblock \url{https://ui.adsabs.harvard.edu/abs/1982A&A...114..245T}

\bibitem[{Tsuge \& Watanabe(2021)}]{tsuge_behavior_2021}
Tsuge, M., \& Watanabe, N. 2021, Acc. Chem. Res., 54, 471.
\newblock \url{https://doi.org/10.1021/acs.accounts.0c00634}

\bibitem[{{Vastel} {et~al.}(2014){Vastel}, {Ceccarelli}, {Lefloch}, \&
  {Bachiller}}]{vastel2014}
{Vastel}, C., {Ceccarelli}, C., {Lefloch}, B., \& {Bachiller}, R. 2014, \apjl,
  795, L2.
\newblock \url{https://ui.adsabs.harvard.edu/abs/2014ApJ...795L...2V}

\bibitem[{Wakelam {et~al.}(2017)Wakelam, Loison, Mereau, \&
  Ruaud}]{wakelam_binding_2017}
Wakelam, V., Loison, J.~C., Mereau, R., \& Ruaud, M. 2017, Mol. Astrophys., 6,
  22.
\newblock
  \url{https://www.sciencedirect.com/science/article/pii/S240567581630032X}

\bibitem[{Weigend \& Ahlrichs(2005)}]{weigend_balanced_2005}
Weigend, F., \& Ahlrichs, R. 2005, PCCP, 7, 3297.
\newblock \url{https://pubs.rsc.org/en/content/articlelanding/2005/cp/b508541a}

\bibitem[{Weigend {et~al.}(2002)Weigend, K{\"o}hn, \&
  H{\"a}ttig}]{weigend_efficient_2002}
Weigend, F., K{\"o}hn, A., \& H{\"a}ttig, C. 2002, \jcp, 116, 3175.
\newblock \url{https://aip.scitation.org/doi/abs/10.1063/1.1445115}

\bibitem[{Woon(2021)}]{woon_quantum_2021}
Woon, D.~E. 2021, Acc. Chem. Res., 54, 490.
\newblock \url{https://doi.org/10.1021/acs.accounts.0c00717}

\bibitem[{{Yang} {et~al.}(2022){Yang}, {Green}, {Pontoppidan}, {Bergner},
  {Cleeves}, {Evans}, {Garrod}, {Jin}, {Kim}, {Kim}, {Lee}, {Sakai},
  {Shingledecker}, {Shope}, {Tobin}, \& {van Dishoeck}}]{Yang2022-JWSTices}
{Yang}, Y.-L., {Green}, J.~D., {Pontoppidan}, K.~M., {et~al.} 2022, \apjl, 941,
  L13.
\newblock \url{https://ui.adsabs.harvard.edu/abs/2022ApJ...941L..13Y}

\bibitem[{{Zamirri} {et~al.}(2018){Zamirri}, {Casassa}, {Rimola},
  {Segado-Centellas}, {Ceccarelli}, \& {Ugliengo}}]{Zamirri2018}
{Zamirri}, L., {Casassa}, S., {Rimola}, A., {et~al.} 2018, \mnras, 480, 1427

\bibitem[{{Zamirri} {et~al.}(2017){Zamirri}, {Corno}, {Rimola}, \&
  {Ugliengo}}]{Zamirri2017}
{Zamirri}, L., {Corno}, M., {Rimola}, A., \& {Ugliengo}, P. 2017, ACS Earth
  Space Chem., 1, 384

\bibitem[{{Zamirri} {et~al.}(2019){Zamirri}, {Ugliengo}, {Ceccarelli}, \&
  {Rimola}}]{Zamirri2019}
{Zamirri}, L., {Ugliengo}, P., {Ceccarelli}, C., \& {Rimola}, A. 2019, ACS
  Earth Space Chem., 3, 1499

\bibitem[{{Zmuidzinas} {et~al.}(1988){Zmuidzinas}, {Betz}, {Boreiko}, \&
  {Goldhaber}}]{Zmuidzinas1988}
{Zmuidzinas}, J., {Betz}, A.~L., {Boreiko}, R.~T., \& {Goldhaber}, D.~M. 1988,
  \apj, 335, 774.
\newblock \url{https://ui.adsabs.harvard.edu/abs/1988ApJ...335..774Z}

\bibitem[{Ãsgeirsson {et~al.}(2021)Ãsgeirsson, Birgisson, Bjornsson, Becker,
  Neese, Riplinger, \& JÃ³nsson}]{asgeirsson_nudged_2021}
Ãsgeirsson, V., Birgisson, B.~O., Bjornsson, R., {et~al.} 2021, JCTC, 17,
  4929.
\newblock \url{https://doi.org/10.1021/acs.jctc.1c00462}

\end{thebibliography}
\bibliographystyle{aasjournal}



\end{document}